\documentclass[11pt, a4paper]{article}
\usepackage{jcappub, physunits, slashed}
\usepackage{setspace, caption, subcaption}
\usepackage{nccmath, physics, tensor}
\usepackage{xparse, mathtools, aas_macros}
\usepackage[toc, page]{appendix}
\usepackage[mathscr]{euscript}
\usepackage{tikz}
\usepackage[compat=1.0.0]{tikz-feynman}
\usepackage{tikzfeynman}

\newcommand{\amp}{\mathcal{A}}
\newcommand{\ham}{\mathcal{H}}
\newcommand{\cphi}{\varphi}
\newcommand{\ep}{\epsilon}
\newcommand{\kap}{\kappa}
\newcommand{\mcal}{\mathcal{M}}
\newcommand{\cd}{\mathcal{D}}
\newcommand{\normal}{\mathcal{N}}
\newcommand{\tord}{\mathcal{T}}
\newcommand{\pow}{\mathcal{P}}
\newcommand{\coll}{\mathcal{C}}
\newcommand{\bp}{\vb{p}}
\newcommand{\br}{\vb{r}}
\newcommand{\bk}{\vb{k}}
\newcommand{\bv}{\vb{v}}

\title{An SZ-Like Effect on Cosmological Gravitational Wave Backgrounds}
\author[a]{Tatsuya Daniel,} 
\author[b,1]{Marcell Howard,\note{Corresponding author.}}
\author[c]{and Morgane K\"onig}

\affiliation[a]{Brown Theoretical Physics Center, Department of Physics, Brown University, Providence, Rhode Island 02912, USA}
\affiliation[b]{Pittsburgh Particle Physics, Astrophysics, and Cosmology Center, Department of Physics and Astronomy, University of Pittsburgh, Pittsburgh, Pennsylvania, 15260, USA}
\affiliation[c]{Department of Physics and Astronomy, Dartmouth College, Hanover, NH 03755, USA,\\ Laboratory of Nuclear Science and Center for Theoretical Physics, Massachusetts Institute of Technology, Cambridge, Massachusetts, 02139, USA}

\emailAdd{tatsuya\_daniel@brown.edu}
\emailAdd{mah455@pitt.edu}
\emailAdd{mkonig@mit.edu}

\abstract{Cosmological gravitational wave backgrounds (CGWBs) are the conglomeration of unresolved gravitational wave signals from early Universe sources, which make them a promising tool for cosmologists. Because gravitons decouple from the cosmic plasma early on, one can consider interactions between gravitons and any particle species that were present in the very early Universe. We show that analogous to the cosmic microwave background, elastic scattering on any cosmological background will induce small distortions in its energy density spectrum. We then quantify the magnitude of these spin-dependent spectral distortions when attributed to the dark matter in the early Universe. Lastly, we give estimates for potentially measurable distortions on CGWBs due to gravitational scattering by primordial black holes.}


\begin{document}
\maketitle

\section{Introduction}

The era of gravitational wave astrophysics and cosmology is upon us. Detections of gravitational waves (GWs) represent the newest window into the inner workings of the Universe~\cite{Abbott16, Abbott_16, Abbott17}. Furthermore, the detection of a background of GWs, i.e. the collection of all unresolved GW signals from either cosmological or astrophysical sources, would be a rich pool of information potentially answering questions about inflation, quantum gravity, and many more~\cite{Phinney01, Caprini18, Bartolo22, Caldwell22, Das22, Renzini22}.

Given the recent discovery of a stochastic gravitational wave background of either astrophysical or cosmological origin by the North American Nanohertz Observatory for Gravitational Waves via the use of pulsar timing arrays \cite{Agazie23, Antoniadis23a, Reardon23, Zic23}, one can ask what sort of information can be teased from the background itself. To this end, one can look to its electromagnetic counterpart for inspiration. Due to the geometric optics approximation, which enables sufficiently low-energy gravitational radiation to have a wave-like behavior, GWs can be very well described by formalism derived for classical scattering of electromagnetic waves. Thus, many of the arguments on the topic of electromagnetic waves can be adapted to gravitational phenomena because many of those lines of reasoning can be generalized for any massless particle that follows boson statistics. As a result, we can look to the cosmic microwave background (CMB) as a guide for potential signals that could show up on a GW background. One plausible area to look for analogous behavior between the CMB and a cosmological gravitational wave background (CGWB) is through imprints from the scattering of particles off CGWBs, akin to the Sunyaev-Zeldovich effect on the CMB.

The Sunyaev-Zeldovich (SZ) effect \cite{Sunyaev68, Sunyaev69, Sunyaev70, Sunyaev80, Hu95, Birkinshaw99} has become an invaluable tool in astrophysics and cosmology over the past decades. This effect corresponds to the small distortion of the intensity spectrum of CMB photons caused by elastic scattering between the CMB photons and electrons occupying hot clouds of molecular gas. This interaction has the effect of preferentially boosting the energy of photons, which induces the presence of hot and cold spots in the CMB. The dip in the intensity typically happens at frequencies $\nu \lesssim 218 \Hz[G]$ \cite{Carlstrom02}, and the brightness is increased for higher frequencies. These distortions are well described by the Kompaneets equation \cite{Kompaneets57}.

In this manuscript, we are interested in investigating the analogous effect for CGWBs by studying the scattering of primordial gravitons and a toy model of thermally produced fundamental particle DM in the very early Universe with a mass of $m = 10\eV[T]$ and  of spin-$s$ where $s \in \{0, \frac{1}{2}, 1\}$. The optical depth of scattering for these candidates in particular and any particles with a mass below the Planck scale is tiny. However, even if the distortion from these particles is observationally immeasurable, the optical depth of scattering for stars and primordial black holes from Compton scattering is $\tau_{\rm stars} = 10^{-8}$ and $\tau_{\rm PBH} = 10^{-5}$ for $M = 10^{-12} M_\odot$ \cite{Cusin19}. Thus, developing the formalism with this toy model will be useful for its application to more observationally interesting massive objects.

The paper is organized as follows: in Section~\ref{CGWB} we provide a brief review of the formalism of inflationary GWs. In Section~\ref{dhk eff} we derive the gravitational analog to the SZ effect on cosmological backgrounds. In Section~\ref{dark matter} we provide the Beyond the Standard Model (BSM) models that could make an imprint on cosmological backgrounds, and in Section~\ref{discuss} we conclude with some remarks and directions for future work.

\paragraph{Conventions}

We use the mostly plus metric signature, i.e. $\eta_{\mu\nu} = (-,+,+,+)$, and units where $c = \hbar = k_B = 1$. The reduced four-dimensional Planck mass is $M_{\rm Pl} = (8\pi G)^{-1/2} \approx 2.43 \times 10^{18} \eV[G]$ and the Laplace operator is defined to be $\nabla^2 = \partial_i \partial^i$. We use boldface letters $\br$ to indicate 3-vectors and $p$ and $k$ to denote (momentum) 4-vectors. Conventions for the curvature tensors, covariant and Lie derivatives are all taken from Carroll \cite{Carroll04}. We closely follow the work done in \cite{Oliveira21} while also elucidating on certain details left implicit. All values for cosmological parameters are taken from \cite{Planck20}. 

\section{Lightning Review of Inflationary Gravitational Waves} \label{CGWB}

Let us give a brief review of primordial GWs and their associated background. A more complete review of the CGWB is given in Appendix \ref{CGWB form}. Consider linearized perturbations around the conformal Friedmann–Lemaître–Robertson–Walker spacetime in the synchronous gauge
\begin{equation}
	\dd{s^2} = a^2(\eta)[-\dd{\eta^2} + (\delta_{ij} + h^{TT}_{ij})\dd{r^i}\dd{r^j}],
\end{equation}
where $a(\eta)$ is the scale factor as a function of conformal time $\eta$, $\delta_{ij}$ is the Kronecker delta and $h^{TT}_{ij}(\br,\eta)$ is the transverse and traceless metric perturbation as a function of co-moving position $\br$ and the conformal time. The equations of motion for the metric perturbation where the source has been turned off is
\begin{equation}
    h''^{TT}_{ij}(\br,\eta) + 2\ham h'^{TT}_{ij}(\br,\eta) - \nabla^2h^{TT}_{ij}(\br,\eta) = 0,
\end{equation}
where primes denote differentiation with respect to conformal time and $\ham \equiv a'/a$ is the conformal Hubble parameter. Moving to Fourier space while expanding the perturbation in the basis of polarization tensors, which are defined to be
\begin{equation}
    e^+_{ij} = \mqty(\pmat{3})_{ij}, \hspace{0.5cm} e^\times_{ij} = \mqty(\pmat{1})_{ij},
\end{equation}
yielding 
\begin{equation}
    \tilde{h}''_A + 2\ham\tilde{h}'_A + k^2\tilde{h}_A = 0,
\end{equation}
where $\tilde{h}^{TT}_{ij}(\bk,\eta) = \tilde{h}_A(\bk,\eta)e^A_{ij}(\bk)$ with co-moving wave vector $\bk$. In the low-energy limit (i.e. far away from the Planck scale $M_{\rm Pl}$), we can conceptualize the Fourier modes of gravitational waves as being the quanta of the perturbation in the metric tensor. The primordial power spectrum can generally parameterized\footnote{This parameterization is particularly relevant for inflationary models. See Sec.~\ref{dhk eff} for a list of GW production mechanisms.} by the following
\begin{equation}
	\pow(k) = A_T(k_*)\qty(\frac{k}{k_*})^{n_T},
\end{equation}
where $k_*$ is a pivot scale, $A_T$ is the amplitude of the spectrum, and $n_T$ is the spectral index which describes the tilt of the spectrum and is related to the ratio of the amplitudes of the scalar and tensor power spectrum $r$ by $n_T = -r/8$. Modern estimates for these values are present in Ref. \cite{Planck20X}, which found the amplitude of the scalar power spectrum to be $A_s = 2.1\times 10^{-9}$. Ref.~\cite{Bicep22} constrained the tensor-to-scalar ratio parameter to be $r < 0.03$. Therefore, present-day constraints on the amplitude of the tensor power spectrum and the tensor spectral index are $A_T = rA_s \lesssim 6.3 \times 10^{-11}$ and $n_T \lesssim -0.00375$. A conventional choice for the pivot scale is $k_* = 0.002\pc[M]^{-1}$ \cite{Maggiore07, Maggiore18}. The above quantity is only good for the \emph{primordial} gravitational wave background and is not the background that would be measured today.

We can now introduce the observable of interest: the fractional energy density spectrum defined by $\rho_c\Omega_{\rm GW}(\nu) = \dv*{\rho_{\rm GW}}{\log\nu}$ with $\rho_c$ being the critical density. The energy density can be further related to the occupation number of gravitons via
\begin{equation}
    \Omega_{\rm GW}(\nu) = \frac{8\pi G}{3}\frac{(4\pi)^2}{H^2_0}\nu^4n(\nu),
\end{equation}
where $H_0$ is the Hubble parameter evaluated today. As alluded to earlier, the energy density spectrum as measured today is related to the primordial energy spectrum via the use of transfer functions.\footnote{See refs. \cite{Dodelson03, Kuroyanagi14, Maggiore18} for models for the transfer function.} The energy density of gravitational waves can now be related to the primordial power spectrum by making use of the relation $\Omega_{\rm GW}(\nu) = \abs{\tord_{\rm GW}(\nu)}^2\Omega^P_{\rm GW}(\nu,z)$, where $\Omega^P_{\rm GW}(\nu,z)$ is the primordial energy density spectrum. Thus, the present-day energy density spectrum relates to the primordial log-space power spectrum by
\begin{equation}
	\Omega_{\rm GW}(\nu) = \frac{\pi^2}{3H_0^2}\nu^2\abs{\tord_{\rm GW}(\nu)}^2\pow(\nu).
\end{equation}

\section{The Gravitational Kompaneets Equation} \label{dhk eff}

Here we derive the gravitational analog to the Kompaneets equation. Recall that the SZ effect on the CMB induces a shift in the intensity spectrum of CMB photons. Similarly, unperturbed gravitons propagating through regions of high density of any BSM model will collide, resulting in a shift in CGWBs' energy density spectrum.

Since we are working in direct analogy to the original works of Sunyaev and Zeldovich, we shall be focusing on graviton-matter interactions where the matter field is both non-relativistic and is in thermal equilibrium. Previous work has been done investigating the interactions between CGWBs and matter in the matter- and dark-energy-dominated Universe \cite{Garoffolo22}; thus, we will be working exclusively in the radiation-dominated Universe (although many of our results can be applied to both the matter- and dark-energy-dominated Universe). Additionally, the radiation-dominated era provides a few other benefits: (1) the production of primordial GWs, which would decouple from the cosmic plasma much earlier than any other particle species and thus would be able to scatter with Standard Model and BSM particles, (2) the largest signal on CGWBs is made at these high redshifts, as can be seen in Figure~\ref{opt_depth}, which would allow us to look at the most optimistic scenario for observing this signal, and (3) next-generation GW detectors could, in principle, have the sensitivities necessary to observe GW frequencies that were emitted so early on \cite{Ricciardone17}.

Many of the arguments that have been used in the case of electrons scattering off the CMB will carry over to gravitons and cosmological gravitational backgrounds. However, there are a few crucial differences that arise when one is trying to track the thermal histories of the respective backgrounds. Most importantly, gravitons are expected to decouple from the cosmic plasma at about the Planck time $t_{\rm Pl} \simeq 5.39 \times 10^{-44}\s$ \cite{Kolb90} after the initial expansion from the classical Big Bang singularity. This difference is felt most strongly in the fact that the initial intensity spectrum of the background will not be set by being in local thermodynamic equilibrium with any of the other species in the Standard Model of particle physics.\footnote{It must be stated that were it not for the initial decoupling, gravitons would also follow a Planck distribution \cite{Weinberg72, Rothman06}, in which case gravitons and photons would be identical in all but one case.} Thus, we are unable to describe the intensity of cosmological backgrounds of cosmological origin with a Bose-Einstein distribution.

This immediately raises the question of what sets the initial intensity spectrum, the answer to which is heavily dependent on the source of the primordial gravitational wave background. One source where the primordial power spectrum (to leading order) is known is inflation. Indeed, for a Hubble parameter $H$, the nearly scale-invariant power spectrum for scalar and tensor perturbations at the time of inflation to leading order is given by
\begin{equation}
    \pow(k) \simeq \frac{16}{\pi}\qty(\frac{H}{M_{\rm Pl}})^2,
\end{equation}
which can be attributed to the power spectrum ``freezing” when the physical wavelength is greater than the Hubble radius. 

Using inflation as our GW factory comes with a few advantages. While there are a plethora of different sources for an early universe CGWB ranging from phase transitions \cite{Kosowsky92a, Kosowsky92b, Kosowsky93, Kamionkowski94, Kosowsky01, Kosowsky02, Gogoberidze07, Caligiuri15, RoperPol20}, topological defects \cite{Jones-Smith08, Fenu09}, reheating \cite{Dufaux07, Amin15, Allahverdi20}, primordial black holes \cite{Domenech21}, particle exchange \cite{Ghigleri15, Ghiglieri20, Ringwald21, Ghiglieri22, Muia23} and many more, primordial GWs from inflation \cite{Guzzetti16, Santos22} are widely seen as being the least speculative \cite{Grishchuk00, Moore14}. Additionally, inflation-sourced GWs would have been produced early in the Universe's history, where it could have in principle been able to interact with every particle that froze out of the cosmic plasma.

We proceed with the derivation while also noting that despite looking at DM, many of the assumptions and arguments that we make apply more generally to any non-relativistic particle at thermal equilibrium that interacts with gravity. We start from DM occupying a thermal bath\footnote{Unitarity bounds on the mass of a fundamental dark matter particle \cite{Griest90} constrains our analysis to consider masses $m \le 30\eV[T]$.} at temperature $T$ with mass $m$, and call the joint occupation number of gravitons and DM $n(\bk,\bp;t)$. In order to track the statistical evolution of both the gravitons and DM in the early Universe, we need the covariant Boltzmann equation
\begin{equation}
    \dv{n}{\lambda} = \pdv{n}{x^\mu}\dv{x^\mu}{\lambda} + \pdv{n}{p^\mu}\dv{p^\mu}{\lambda} = \coll[n],
\end{equation}
where $\lambda$ is an arbitrary affine parameter and $\coll[n]$ refers to the collision term. Since the Universe is statistically homogeneous and isotropic\footnote{We neglect any anisotropies that may already be present on the CGWB. See Refs. \cite{Alba16, Contaldi17, Bartolo19} for additional sources and imprints of anisotropies.}, we can assume that the occupation index will similarly be homogeneous, and thus the $\partial_i n$ and $\partial_{p_i}n$ terms drop out. This reduces the left-hand side of the Boltzmann equation to the much friendlier form
\begin{equation}
    \pdv{n(\bk,t)}{t} - H\omega\pdv{n}{\omega} = \coll[n],
\end{equation}
where $\bp$ is the co-moving incoming dark matter 3-momentum with outgoing co-moving momentum $p'^\mu = \qty(E',\bp')$, and $\bk = \omega\vu{n}$ is the incoming co-moving graviton momentum with outgoing co-moving momentum $k'^\mu = \qty(\omega',\bk')$. Similar to photons, there are only two effects that will affect the evolution of the graviton occupation number: stimulated emission,\footnote{See \cite{Weinberg72, Boughn06, Rothman06, Pignol07, Hu21} on how stimulated emission of gravitational radiation has been previously explored.} which occurs when the frequency of the incident graviton matches the difference in the energy between two quantum states, resulting in the emission of a graviton whose frequency, polarization, and direction of propagation are identical and therefore only affects the graviton occupation number $n(\bk,t)$, and Compton scattering. 

Because stimulated emission results in the creation of a new graviton with identical frequency and polarization, the occupation number is altered to be $n(\bk,t)\rightarrow 1 + n(\bk,t)$. If we assume that the collisions between DM and gravitons are entirely uncorrelated, then the collision term becomes what Boltzmann described as a \emph{Stosszahlansatz} \cite{Hees16}. This term is also known as the ``molecular chaos assumption" where the collision term is regarded as being a product of one-particle distributions integrated over the momenta of the particles. Working in a locally Minkowski spacetime for the collision term \cite{Bernstein88, Hu95}, this results in the following integro-differential equation that we will call the Boltzmann-Master equation for both gravitons and DM
\begin{equation}
    \begin{split}
        \pdv{n(\bk;t)}{t} - H\omega\pdv{n}{\omega} = \int\dd[3]{\bp}\int\dd[3]{\bp'}\int\dd[3]{\bk'}&\left[n(\bp',\bk';t)w(p',k'\rightarrow p,k)(1 + n(\bk,t))\right. \\ &-\left. n(\bp,\bk;t)w(p,k\rightarrow p',k')(1 + n(\bk',t))\right],
    \end{split}
\end{equation}
$w(p,k\rightarrow p',k')$ characterizes the rate of transition between $(p,k) \leftrightarrow (p',k')$. Written in this way, $w(p,k\rightarrow p',k')$ is often called the non-covariant transition rate. It can be related to the covariant transition rate $W(p,k\rightarrow p',k')$ \cite{Hees16} by 
\begin{equation}
    w(p,k\rightarrow p',k')\dd[3]{\bp'}\dd[3]{\bk'} = \frac{\dd[3]{\bp'}\dd[3]{\bk'}}{\omega E\omega'E'}W(p,k\rightarrow p',k') = \qty(1 - \bv\cdot\vu{n})\dv{\sigma_s(\bp,\bk,\bk')}{\Omega}\dd{\Omega},
\end{equation}
where $(1 - \bv\cdot\vu{n})$ is the M$\o$ller velocity, which in the non-relativistic limit is the relative velocity between the gravitons and dark matter, and $\dv*{\sigma_s}{\Omega}$ is the differential gravitational Compton cross section for an arbitrary spin-$s$ massive target. The energy shift of the graviton is given by the (dimensionless) Compton scattering equation
\begin{equation}
    \Delta(\omega,\bp,\vu{n}',\vu{n}) \equiv \frac{\omega' - \omega}{T} = \frac{\omega}{T}\frac{\bp\cdot(\vu{n}' - \vu{n}) - \omega(1 - \vu{n}\cdot\vu{n}')}{E - \bp\cdot\vu{n}' + \omega(1 - \vu{n}\cdot\vu{n}')}.
\end{equation}
We make the assumption that the DM gas evolves on a much shorter timescale than the gravitons. Hence the joint occupation number factorizes into the graviton occupation number and the DM momentum distribution $n(\bp,\bk;t) = \normal(\bp)n(\bk,t)$, where $\normal(\bp)$ is the three-dimensional momentum distribution of DM assuming isotropic conditions. The Boltzmann equation becomes
\begin{equation}
    \begin{split}
        \pdv{n(\bk,t)}{t} - H\omega\pdv{n}{\omega} = \int\dd[3]{\bp}\int\dd[3]{\bp'}\int\dd[3]{\bk'}&\left[\normal(\bp')n(\bk',t)w(p',k'\rightarrow p,k)(1 + n(\bk,t))\right.\\ &-\left. \normal(\bp)n(\bk,t)w(p,k\rightarrow p',k')\qty(1 + n(\bk',t))\right].
    \end{split}
\end{equation}
\cite{Hu95} showed that 
\begin{equation}
    H\omega\pdv{n}{\omega} = \omega\pdv{n}{\omega}\pdv{t}\ln(\frac{T}{T_g(1 + z)}),
\end{equation}
where $T_g$ is the temperature of the gravitons. Additionally, since the DM as well as the graviton temperatures are tightly coupled to the photon temperature, \cite{Hu95} showed that $T \propto (1 + z)$ and can thus be dropped entirely. Given that DM will be moving non-relativistically after it thaws out, its distribution, regardless if its true nature is fermionic or bosonic, can be described by a simple Maxwell-Boltzmann distribution and we can therefore write
\begin{equation}
    \normal(\bp)\dd[3]{\bp} = \normal_{\rm eq}(|\bp|)\dd[3]{\bp} = \frac{n_\chi}{(2\pi mT)^{3/2}}\exp(-\frac{\bp^2}{2mT})\dd[3]{\bp},
\end{equation}
where $n_{\chi}$ is the physical number density of dark matter given by $n_{\chi} = \rho_\chi/m$ with $\rho_\chi \simeq 0.3\eV[G]/\cm^3$ \cite{TASI19, Baxter21}. We also note that conservation of energy enforces 
\begin{equation}
    E' = E - T\Delta,\hspace{0.5cm} \normal_{\rm eq}(\bp') = \normal_{\rm eq}(\bp)\qty(1 + \Delta + \frac{\Delta^2}{2}).
\end{equation}
Now we assume that even though the DM have long thawed out (become non-relativistic, i.e. $|\bp|$ and $T\ll m$ where $T$ is the temperature of the DM) well into the recombination era, we also have soft gravitons, i.e. $\omega \ll m$ but $\omega\sim T$. 

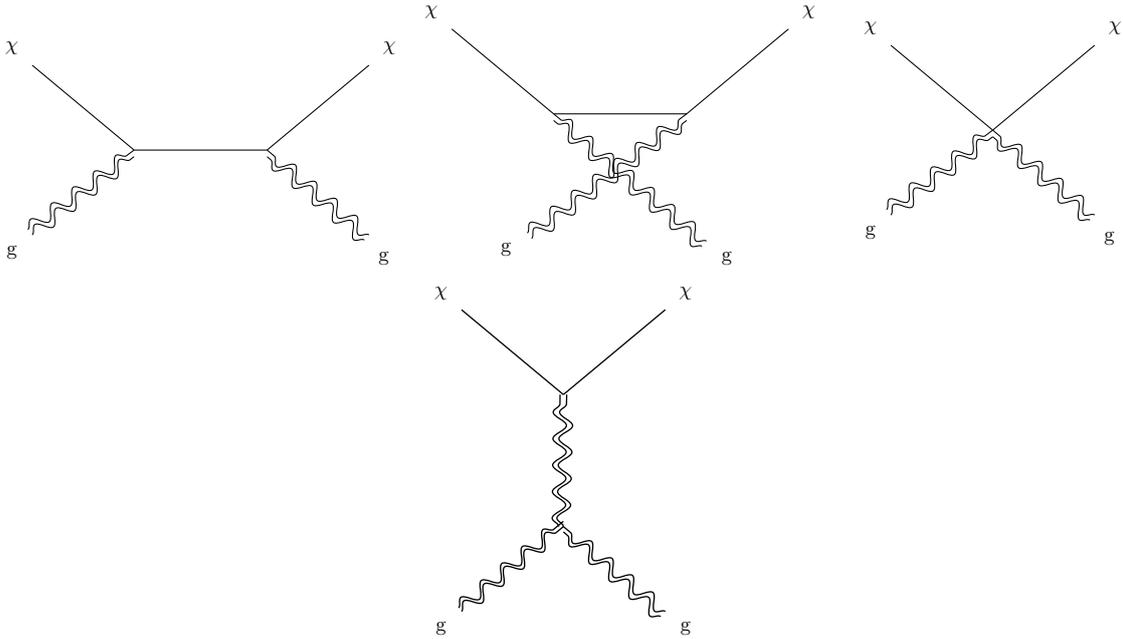
\begin{figure}[t!]
    \centering
    \raisebox{-0.55\height}{
    \begin{tikzpicture}
    \begin{feynman}[line width=.5 pt, scale=1.75]
	\draw[fermionnoarrow] (0:1)--(0,0);
	\draw[fermionnoarrow] (140:1)--(0,0);
	\draw[vector] (217:1)--(0,0);
	\draw[vector] (220:1)--(0,-.05);
    \node at (220:1.2) {\scalebox{.7}{g}};
    \node at (140:1.2) {\scalebox{.7}{$\chi$}};
 \begin{scope}[shift={(0:1)}]
    \draw[fermionnoarrow] (40:1)--(0,0);
    \draw[vector] (-40:1)--(0,0);
    \draw[vector] (-43:1)--(0,-.05);
    \node at (-43:1.2) {\scalebox{.7}{g}};
    \node at (40:1.2) {\scalebox{.7}{$\chi$}};   
 \end{scope} 
 \end{feynman}
 \end{tikzpicture}

  \begin{tikzpicture}
    \begin{feynman}[line width=.5 pt, scale=1.75]
	\draw[fermionnoarrow] (0:1)--(0,0);
	\draw[fermionnoarrow] (140:1)--(0,0);
    \draw[vector] (-40:1.5)--(0,0);
    \draw[vector] (-42:1.5)--(0,-.05);
    \node at (-40:1.7) {\scalebox{.7}{g}};
    \node at (140:1.2) {\scalebox{.7}{$\chi$}};
 \begin{scope}[shift={(0:1)}]
    \draw[fermionnoarrow] (40:1)--(0,0);
    \draw[vector] (217:1.5)--(0,0);
	\draw[vector] (219:1.5)--(0,-.05);
    \node at (217:1.7) {\scalebox{.7}{g}};
    \node at (40:1.2) {\scalebox{.7}{$\chi$}};   
 \end{scope} 
 \end{feynman}
 \end{tikzpicture}
    }
           \centering
           \raisebox{-0.55\height}{
    \begin{tikzpicture}
    \begin{feynman}[line width=.5 pt, scale=1.75]
	\draw[fermionnoarrow] (140:1)--(0,0);
	\draw[vector] (217:1)--(0,0);
	\draw[vector] (220:1)--(0,-.05);
    \node at (220:1.2) {\scalebox{.7}{g}};
    \draw[vector] (-40:1)--(0,0);
    \draw[vector] (-43:1)--(0,-.05);
    \draw[fermionnoarrow] (40:1)--(0,0);
	\node at (-43:1.2) {\scalebox{.7}{g}};
    \node at (40:1.2) {\scalebox{.7}{$\chi$}};
	\node at (140:1.2) {\scalebox{.7}{$ \chi$}};
 \end{feynman}
\end{tikzpicture}
    }
           \raisebox{-0.55\height}{
     \begin{tikzpicture}[line width=.5 pt, scale=1.75]
    \draw[fermionnoarrow] (140:1)--(0,0);
	\draw[fermionnoarrow] (40:1)--(0,0);
	\draw[vector] (-90:1)--(0.025,0);
	\draw[vector] (-92:1)--(-0.025,0);
	\node at (140:1.2) {\scalebox{.7}{$\chi$}};
	\node at (40:1.2) {\scalebox{.7}{$\chi$}};
    \begin{scope}[shift={(-90:1)}]
    \draw[vector] (-40:1)--(0,0);
    \draw[vector] (-43:1)--(0,-.04);
    \draw[vector] (220:1)--(0,0);
    \draw[vector] (218:1)--(0,0.04);
    \node at (-40:1.2) {\scalebox{.7}{g}};
    \node at (220:1.2) {\scalebox{.7}{g}};
    \end{scope}
 \end{tikzpicture}
    }
    \caption{The Feynman diagrams relevant for tree-level gravitational Compton scattering. }
    \label{feynman}
\end{figure}

Noticing that the $\Delta$ dependence on the ratio $\omega/m$ will ensure that it is small, we can simplify the integrand of the Boltzmann equation by first making a change of variables $\omega\rightarrow x = \omega/T$, $\omega'\rightarrow x' = \omega'/T$, and expanding to quadratic order in $\Delta$
\begin{equation}
    n(x',t) = n(x + \Delta, t) = n(x, t) + \Delta\pdv{n}{x} + \dfrac{\Delta^2}{2}\pdv[2]{n}{x},
\end{equation}
we can similarly expand the function $1 + n(x',t)$ which leads into an equation that is almost identical to the Kompaneets equation
\begin{equation} \label{grav_komp}
    \pdv{n}{t} = \frac{J_2(x,\lambda;s)}{2}\pdv{x}\qty[\pdv{n}{x} + n(1 + n)] + \qty(J_1(x,\lambda;s) + \frac{J_2(x,\lambda;s)}{2})\qty[\pdv{n}{x} + n(1 + n)],
\end{equation}
where the quantities $J_\ell(x,\lambda;s)$ have a more complicated dependence on the dimensionless frequency
\begin{equation} \label{jell}
    J_{\ell}(x,\lambda;s) = 2\pi\int_{\theta_{\min}(\lambda)}^{\theta_{\max}}\sin\theta\dd{\theta}\int\dd[3]{\bp}\qty(1 - \frac{\bp}{m}\cdot\vu{n})\dv{\sigma_s(\bp,x,\theta)}{\theta}\normal_{\rm eq}(\bp)\Delta^{\ell}(x,\theta),
\end{equation}
where we have to regularize the $\theta$-integral by introducing cutoff angles $\theta_{\max},\ \theta_{\min}(\lambda)$. This is the second major difference between the gravitational and electromagnetic case and is also the same type of divergence one sees in the problem of classical Coulomb scattering. One can trace back the origin of this divergence to the presence of a pole in the $t$-channel process (notice the last diagram in Figure~\ref{feynman} with a self-interaction term for the gravitons). This is referred in the literature  as the forward scattering limit where $t \equiv -(k - k')^2 \sim 0$ when $\theta \rightarrow 0$. Essentially, a DM particle can interact with a real graviton, emit a virtual graviton, and still propagate with practically no change in its momentum (i.e. momentum transfer of zero). 

This divergence can also be attributed to the fact that we have essentially taken the non-relativistic limit in the computation of the differential cross section. As $c$ tends towards infinity, signals propagate instantaneously and therefore there is no long-ranged screening. Consequently, the cross section becomes infinitely large since contributions to scattering from particles located infinitely far away in all directions must be taken into account. To get around this issue, we implement the cutoff suggested in \cite{Cusin19} where the lower cutoff angle is set at the geometric optics limit with cutoff\footnote{Other treatments for dealing with angular singularities for long-range scattering can be found in \cite{Villani02, Alexandre04, He21}.} wavelength $\lambda_{\rm GO}$. In principle, one would have to use the metric of a static point particle. That metric, derived in \cite{Katanaev13}, is
\begin{equation}
    \dd{s^2} = -\qty(\frac{1 - \frac{Gm}{2r}}{1 + \frac{Gm}{2r}})^2\dd{t^2} + \qty(1 + \frac{Gm}{2r})^4\qty[\dd{r^2} + r^2\dd{\theta^2} + r^2\sin[2](\theta)\dd{\cphi^2}],
\end{equation}
which is the Schwarzschild metric in isotropic coordinates\footnote{The metric for point particles of spin 0, 1/2, and 1 was also worked out in \cite{Donoghue02, Holstein_06} and recovers the metric in \cite{Katanaev13} at leading order in $r_s$.} with a Kretschmann scalar given by
\begin{equation}\label{kretsch}
    \bar{R}^{\mu\nu\lambda\rho}\bar{R}_{\mu\nu\lambda\rho} \equiv K(r) = \frac{12r_s^2r^6}{(r + \frac{r_s}{4})^{12}} \ll \frac{1}{\lambda^4_{\rm GO}},
\end{equation}
with $r_s = 2GM$ being the Schwarzschild radius and $\lambda_{\rm GO}$ the wavelength\footnote{We label this wavelength by $\lambda_{\rm GO}$ for geometric optics in order to reinforce the notion that this wavelength is completely independent of the energy of the initial scattering graviton $\omega$.} of an arbitrary GW. However, for a point particle, $r_s$ will be negligible compared to $r$, and therefore the Kretschmann scalar reduces to that of a compact object
\begin{equation}
    K(r) \simeq \frac{12r_s^2}{r^6}.
\end{equation}
One can then use equation (\ref{kretsch}) to define a region $K(R_\lambda)$ for which wave-like effects are relevant, which is given by
\begin{equation}
    R_\lambda = \qty(2\sqrt{3}r_s\lambda^2_{\rm GO})^\frac{1}{3}.
\end{equation}
This region provides a maximum value for the impact parameter $b_{\max}(\lambda) = R_\lambda$. Now arises the question of determining the most appropriate length scale to utilize in this scattering process. \cite{Cusin19, Pizzuti22} places that length scale at the Schwarzschild radius. Since our focus is on the elastic scattering of gravitons and point-like particles, the more pertinent length scale would be the Compton wavelength $\lambda_C = m^{-1}$. This wavelength effectively establishes the physical size of the particle, serving as the radius that the graviton ``sees" when colliding. Indeed, one can see from the ratio of the Schwarzschild radius and Compton wavelength
\begin{equation}\label{schwarzcomp}
    \frac{r_s}{\lambda_C} \sim \frac{m^2}{M^2_{\rm Pl}},
\end{equation}
that as long as we are far away from masses that are at the Planck mass and above, the Compton wavelength scale will dominate. Inversely, for masses above the Planck mass (e.g. stellar mass objects such as black holes and neutron stars), the Schwarzschild radius will be a far more relevant scale. Thus we set the minimum angle to be 
\begin{equation}
    \theta_{\min}(\lambda) = \frac{\lambda_C}{b_{\max}(\lambda)} = \qty(\frac{2Gm}{\lambda_{\rm GO}})^{2/3}\frac{1}{Gm^2}.
\end{equation}
Consequently, one can only see that Eqn (\ref{schwarzcomp}) also defines a range of masses for which processes such as absorption would become relevant as well. Thus we can safely take $\theta_{\max} = \pi$ for all DM masses that are far away from the Planck scale. We must note that because our choice of cutoff angle is wavelength-dependent, the gravitational Kompaneets equation will have an extra redshift dependence that is not present in the original Kompaneets equation.

Going back to Eqn. (\ref{grav_komp}), we recognize that any deviations in the occupation index of gravitons will be small. Thus, we can expect a power law behavior\footnote{This simple power law relation is a consequence of the power spectrum for single-field inflation being nearly scale invariant.} $n(x) = A_T x^\alpha$, which brings the gravitational Kompaneets equation to the form
\begin{equation}
    \begin{split}
        \pdv{n}{t} &= A_T(k_*)\qty[J_1(x,\lambda;s) + \frac{J_2(x,\lambda;s)}{2} + \frac{\alpha(J_1(x,\lambda;s) + J_2(x,\lambda;s))}{x} + \frac{\alpha(\alpha - 1)J_2(x,\lambda;s)}{2x^2}]x^\alpha\\ &+ A_T^2(k_*)\qty(J_1(x,\lambda;s) + \frac{J_2(x,\lambda;s)}{2} + \frac{\alpha J_2(x,\lambda;s)}{x})x^{2\alpha},
    \end{split}
\end{equation}
Hence all of the deviations in the occupation index can be attributed to the specific choices of $J_\ell(x,\lambda;s)$. We note, since $A_T \ll 1$, the $\order{A^2_T}$ terms can be safely discarded. 

\section{Dark Matter Candidates} \label{dark matter}

\subsection{Spin-$s$ Fundamental Particles}
Now that we have established that a similar distortion on the energy density spectrum for cosmological backgrounds can arise as a result of an inverse-Compton scattering event, it is natural to inquire what the insights we can gain from this observation. In the case of the CMB, the SZ effect gives information on certain properties of galaxy clusters (total mass of the cluster, velocity dispersion, etc.). For our purposes, we can use cosmological backgrounds to probe particle interactions between gravitons and BSM particles. We give a particular focus on BSM models that double as DM candidates, as these provide estimates of their number density. In particular, since the distortion on CGWBs depends on the cross section between the graviton and the target particle, DM of differing spins will induce a different distortion on the energy density spectrum. Thus, we shall solve the gravitational Kompaneets equation for DM of varying spin where we shall only assume minimal coupling between DM and gravity.

We start with a DM candidate $\chi$ that couples minimally to gravity with spin-0 whose action is given by
\begin{equation}
    S_{s = 0} = -\frac{1}{2}\int\dd[4]{x}\sqrt{-g}\qty(g^{\mu\nu}\nabla_\mu\chi\nabla_\nu\chi + m^2\chi^2),
\end{equation}
where $\nabla_\mu$ is the covariant derivative with the Levi-Civita connection. The spin-1/2 diffeomorphism-invariant action is given by
\begin{equation}
    S_{s = \frac{1}{2}} = \int\dd[4]{x}\sqrt{-g}\overline{\chi}\qty(i\slashed{\cd} - m)\chi, 
\end{equation}
with $\slashed{\cd} = \gamma^\mu(\partial_\mu + \frac{1}{2}\omega_{\mu ab}\gamma^{ab})$, $\partial_\mu \equiv e^a_\mu\partial_a$, $\{e_\mu^a\}$ are the orthonormal frame fields, $\omega^{ab}_\mu = -e^{b\nu}\nabla_\mu e^a_\nu$ is the spin connection and $\gamma^{ab} = -\frac{i}{4}\comm{\gamma^a}{\gamma^b}$. Here, Latin indices are used to denote the internal flat spacetime while Greek indices represent spacetime coordinates. Lastly, we have the action for a spin-1 massive vector field
\begin{equation}
    S_{s = 1} = \int\dd[4]{x}\sqrt{-g}\qty[-\frac{1}{4}g^{\mu\nu}g^{\lambda\rho}\chi_{\mu\lambda}\chi_{\nu\rho} - \frac{1}{2}m^2g^{\mu\nu}\chi_\mu\chi_\nu],
\end{equation}
where $\chi_{\mu\nu} = \partial_\mu\chi_\nu - \partial_\nu\chi_\mu$. Ordinarily, given some Lagrangian, one would start expanding the metric around some background spacetime $g_{\mu\nu} = g^{(0)}_{\mu\nu} + \kap h_{\mu\nu}$, where $g^0_{\mu\nu}$ is the background metric (for our purposes it suffices for the background to be Minkowski) and $h_{\mu\nu}$ is the perturbation in the metric with coupling constant $\kap = \sqrt{32\pi G}$. We would next have to express the action in terms of this new dynamical field, write out the Feynman rules between the perturbation and the matter fields after choosing a suitable gauge on gravity, and proceed through a series of very long and tedious calculations. Thankfully, we can circumvent this process by utilizing spinor-helicity variables.

The spinor-helicity formalism takes advantage of the anti-commuting properties of spinors by associating the momentum of a given field to a 2- or 4-component Weyl spinor. Massless particles carry with them either a positive or negative helicity which can in turn be mapped to a(n) (anti-)spinor that also carries positive or negative helicity. Due to this mapping, the formalism is particularly well-suited whenever one is interested in the scattering of massless particles \cite{Chung19}; however, there have been developments to extend this formalism to massive particles \cite{Cruz16, Arkani-Hamed17}. We give a more detailed summary in Appendix~\ref{spin-hel} and leave a much more exhaustive introduction to the subject in \cite{Elvang13, Elvang15}, as well as \cite{Burger18}, which is particularly helpful for introducing astrophysicists to the topic within the context of gravitational scattering of massive bodies, the subject of this manuscript. 

In the lab frame, the differential cross section for Compton scattering takes the form
\begin{equation}
    \dv{\sigma_s(\bp,\vu{n},\vu{n}')}{\Omega} = \frac{\sigma_s(\lambda)}{(4\pi)^2}\frac{1}{(E - \bp\cdot\vu{n})^2}\qty(\frac{\omega'(\bp,\vu{n},\vu{n}')}{\omega})^2\abs{\overline{\mcal_{s}}}^2,
\end{equation}
where $\sigma_s(\lambda)$ is the (low energy)\footnote{We define this quantity for each spin in Appendix~\ref{optical_depth}.} rest frame cross section between a graviton and spin-$s$ particle evaluated at the geometric optics wavelength $\lambda_{\rm GO}$ (we choose to suppress the subscripts to make the expressions easier to read), the energy $E$ and momentum $\bp$ belong to the DM with $\omega, \omega'$ satisfying $\bk = \omega\vu{n}$ and $\bk' = \omega'\vu{n}'$, similar to the case of electron-photon scattering, and their ratio is given by the formula from Compton scattering
\begin{equation}
    \frac{\omega'(\bp,\vu{n},\vu{n}')}{\omega} = \frac{E - \bp\cdot\vu{n}}{E - \bp\cdot\vu{n}' + \omega(1 - \vu{n}\cdot\vu{n}')}.
\end{equation}
The gravitational Kompaneets equation clearly depends on the differential cross sections, which imply a new Kompaneets equation for each spin. The modulus-squared amplitude takes the form
\begin{equation}
    \abs{\mcal_{s = 0}}^2 = \frac{\kap^4}{8}\qty[\frac{m^8(1 - \vu{n}\cdot\vu{n}')^2}{(E - \bp\cdot\vu{n})^2(E - \bp\cdot\vu{n}')^2} + \frac{[2(E - \bp\cdot\vu{n})(E - \bp\cdot\vu{n}') - m^2(1 - \vu{n}\cdot\vu{n}')]^4}{(1 - \vu{n}\cdot\vu{n}')^2(E - \bp\cdot\vu{n})^2(E - \bp\cdot\vu{n}')^2}]
\end{equation}
for graviton-scalar scattering,
\begin{equation}
    \begin{split}
        \abs{\mcal_{s = \frac{1}{2}}}^2 =&\  \frac{\kap^4}{4}\left[\frac{m^8(1 - \vu{n}\cdot\vu{n}')^2}{(E - \bp\cdot\vu{n})^2(E - \bp\cdot\vu{n}')^2} + \frac{m^6\omega\omega'(1 - \vu{n}\cdot\vu{n}')^3}{(E - \bp\cdot\vu{n})^2(E - \bp\cdot\vu{n}')^2}\right.\\ &\left.+\ \frac{[2(E - \bp\cdot\vu{n})(E - \bp\cdot\vu{n}') - m^2(1 - \vu{n}\cdot\vu{n}')]^3}{(1 - \vu{n}\cdot\vu{n}')^2(E - \bp\cdot\vu{n})^2(E - \bp\cdot\vu{n}')^2}\times\right.\\&\left.\times\qty(\frac{\omega'}{\omega}(E - \bp\cdot\vu{n}')^2 + \frac{\omega}{\omega'}(E - \bp\cdot\vu{n})^2 - m^2(1 - \vu{n}\cdot\vu{n}'))\right]
	\end{split}
\end{equation}
in the case for graviton-spinor scattering, and finally
\begin{equation}
	\begin{split}
		  \abs{\mcal_{s = 1}}^2 =& \ \frac{\kap^4}{2}\left[\frac{m^8(1 - \vu{n}\cdot\vu{n}')^2}{(E - \bp\cdot\vu{n})^2(E - \bp\cdot\vu{n}')^2} + \frac{4m^4(\omega\omega')^2(1-\vu{n}\cdot\vu{n}')^4}{(E - \bp\cdot\vu{n})^2(E - \bp\cdot\vu{n}')^2}\right.\\ &\left. +\ \frac{2m^6(1 - \vu{n}\cdot\vu{n}')^2(m^2 + 4\omega\omega'(1 - \vu{n}\cdot\vu{n}'))}{(E - \bp\cdot\vu{n})^2(E - \bp\cdot\vu{n}')^2}\right.\\ &\left.+\ \frac{[2(E - \bp\cdot\vu{n})(E - \bp\cdot\vu{n}') - m^2(1 - \vu{n}\cdot\vu{n}')]^2}{(1 - \vu{n}\cdot\vu{n}')^2(E - \bp\cdot\vu{n})^2(E - \bp\cdot\vu{n}')^2}\times\right.\\ &\left.\times\qty(\frac{\omega'}{\omega}(E - \bp\cdot\vu{n}')^2 + \frac{\omega}{\omega'}(E - \bp\cdot\vu{n})^2 - m^2(1 - \vu{n}\cdot\vu{n}'))^2\right]
	\end{split}
\end{equation}
for graviton-massive vector boson scattering with $\abs{\mcal_{s}}^2 = (Gm)^2\abs{\overline{\mcal_{s}}}^2$. One can check that these momentum-dependent differential cross sections, when taking the $\bp\rightarrow0$ limit, recover the cross sections found in \cite{Holstein06, Voronov73, Bohr15}. It should be noted that it is common in the literature to take all the momenta as being outgoing, i.e. conservation of the 4-momentum is written as $p + p' + k + k' = 0$. However, to make contact with transition collisions from the derivation in the previous sections, we have let $p', k'\rightarrow -p', -k'$, which is reflected in the squared amplitudes written down.

Expanding $\qty(1 - \bp\cdot\vu{n}/m)\Delta^\ell\dv*{\sigma_s}{\Omega}$ to second order in $\bp/m$ and doing the Gaussian momentum integrals gives
\begin{equation}
    \begin{split}
        J_1(x,\lambda;0) &= \frac{8\sigma_0n_\chi T}{m}x(4 - x)\qty[2\pi\int^{\beta}_{-1}\frac{(1 - \cos\theta)^4 + (1 + \cos\theta)^4}{1 - \cos\theta}\dd{\cos\theta}],\\ \hspace{0.5cm} J_2(x,\lambda;0) &= \frac{8\sigma_0n_\chi T}{m}2x^2\qty[2\pi\int^{\beta}_{-1}\frac{(1 - \cos\theta)^4 + (1 + \cos\theta)^4}{1 - \cos\theta}\dd{\cos\theta}],
    \end{split}
\end{equation}
where $\beta = \cos((Gm^4\lambda^2_{\rm GO})^{-\frac{1}{3}})$. This particular form is very interesting and is explored in more detail in Appendix~\ref{energy_trans}. For spin-1/2 and spin-1 particles, we get
\begin{equation}
    \begin{split}
        J_1\qty(x,\lambda;\frac{1}{2}) &= \frac{16\sigma_{1/2}n_\chi T}{m}x(4 - x)\qty[2\pi\int^{\beta}_{-1}\frac{(1 + \cos\theta)[(1 - \cos\theta)^4 + (1 + \cos\theta)^3]}{1 - \cos\theta}\dd{\cos\theta}],\\ J_2\qty(x,\lambda;\frac{1}{2}) &= \frac{16\sigma_{1/2}n_\chi T}{m}2x^2 \qty[2\pi\int^{\beta}_{-1}\frac{(1 + \cos\theta)[(1 - \cos\theta)^4 + (1 + \cos\theta)^3]}{1 - \cos\theta}\dd{\cos\theta}],
    \end{split}
\end{equation}
and
\begin{equation}
	\begin{split}
		J_1\qty(x,\lambda;1) &= \frac{32\sigma_1n_\chi T}{m}x(4 - x)\qty[2\pi\int^{\beta}_{-1}\frac{(1 + \cos\theta)^2[3(1 - \cos\theta)^4 + (1 + \cos\theta)^2]}{1 - \cos\theta}\dd{\cos\theta}],\\ J_2\qty(x,\lambda;1) &= \frac{32\sigma_1n_\chi T}{m}2x^2 \qty[2\pi\int^{\beta}_{-1}\frac{(1 + \cos\theta)^2[3(1 - \cos\theta)^4 + (1 + \cos\theta)^2]}{1 - \cos\theta}\dd{\cos\theta}].
	\end{split}
\end{equation}
Defining new variables $\tilde{J}_\ell \equiv  mJ_\ell/(\sigma_{s}Tn_{\chi})$,
we can express the gravitational Kompaneets equation in the following way
\begin{equation} \label{eval_komp}
    \pdv{n}{y} = A_T(k_*)\qty[\tilde{J}_1(x,\lambda;s) + \frac{1}{2}\tilde{J}_2(x,\lambda;s) - \frac{2(\tilde{J}_1(x,\lambda;s) + \tilde{J}_2(x,\lambda;s))}{x} + \frac{3\tilde{J}_2(x,\lambda;s)}{x^2}]x^{-2},
\end{equation}
where we have set $\alpha = -2$ because $n \propto x^{-2}\pow$, and the gravitational Compton-$y$ parameter is defined to be
\begin{equation}
    y = \tau(m, \lambda, z_{\rm scat}, z_O)\frac{T}{m},
\end{equation}
where $\tau(m, \lambda, z_{\rm scat}, z_O)$ is the optical depth for the spin-$s$ DM candidate as a function of the observed redshift $z_O$ and the redshift of scattering $z_{\rm scat}$\footnote{We fix the observed redshift to matter-dark energy equality i.e. $z_{O} = (\Omega_\Lambda/\Omega_M)^{1/3} - 1$.} whose explicit form is given by
\begin{equation}
    \tau(m, \lambda, z_{\rm scat}, z_O) = \int^{\eta_{\rm scat}(z_{\rm scat})}_{\eta_O(z_{O})}\dd{\eta}n_{\chi}(\eta)\sigma_s(m, \lambda, \eta)a(\eta) = \int^{z_{\rm scat}}_{z_O}n_\chi(z)\sigma_s(m,\lambda,z)a(z)\dv{\eta}{z}\dd{z},
\end{equation}
where the $\dv*{\eta}{z}$ is explicitly written to indicate that a change of variables has taken place. The optical depth is most sensitive to the mass of the DM and the cutoff wavelength. Therefore, it would be handy to express its dependence on both quantities
\begin{equation}
    \tau(m,\lambda,z) \equiv \frac{12\pi(G^5m^{14}\lambda^4_{\rm GO})^{1/3}H_0\Omega_\chi}{\Omega_M^{1/2}}f(z) = \frac{m^{14/3}\lambda^{4/3}_{\rm GO}}{\xi}f(z),
\end{equation}
where $f(z)$ is a function that contains all of the redshift dependencies of the optical depth and the constant $\xi$ is given by
\begin{equation}\label{xi_constant}
    \xi = \frac{\Omega_M^{1/2}}{12\pi G^{5/3}H_0\Omega_\chi}.
\end{equation}
\begin{figure}[h!]
  \centering
  \includegraphics[width=12cm]{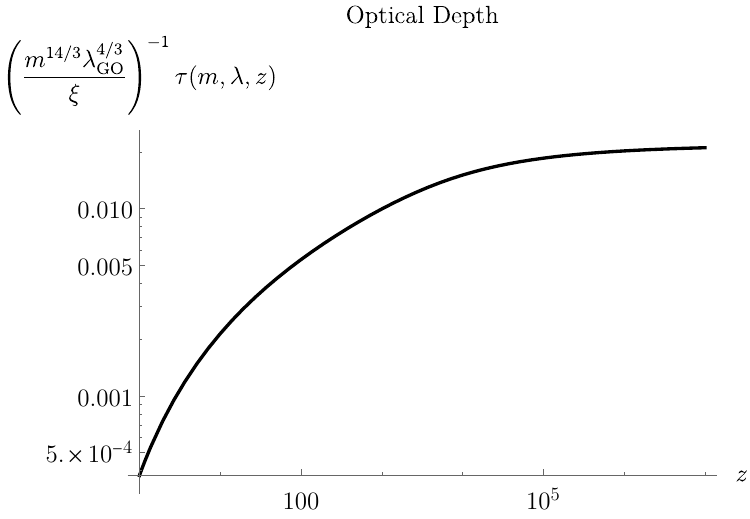}
  \caption{The optical depth as a function of redshift for mass $m$ DM candidate. The constant $\xi$ is given in Eqn. (\ref{xi_constant}).}
  \label{opt_depth}
\end{figure}

In Figure~\ref{opt_depth}, we plot the optical depth as a function of redshift. We display a DM candidate with a mass, $m$, impinging off a graviton with a cutoff wavelength $\lambda_{\rm GO}$. One can choose values that corresponds to where current GW detectors will be the most sensitive \cite{Moore14, Moore15, Martynov16, Robson19, Schmitz21}. This plot can be seen as a representative sample of the shape and behavior for the optical depth evaluated for different spins. 

Going back to equation (\ref{eval_komp}), we can estimate the primordial amplitude for CGWBs to be $A_T \sim 16H^2/(\pi M^2_{\rm Pl})$, and by again approximating $\pdv*{n}{y} \simeq \Delta n/y$, we can write
\begin{equation}
    \Delta n(x,z;s) = y\frac{16}{\pi}\qty(\frac{H}{M_{\rm Pl}})^2\qty[\tilde{J}_1(x,\lambda;s) + \frac{1}{2}\tilde{J}_2(x,\lambda;s) - \frac{2(\tilde{J}_1(x,\lambda;s) + \tilde{J}_2(x,\lambda;s))}{x} + \frac{3\tilde{J}_2(x,\lambda;s)}{x^2}]x^{-2},
\end{equation}
where we can evaluate the temperature $T$ to be $T_0(1 + z)^2$ with $T_0$ being the present temperature of the CMB and $z$ being the redshift at which DM thaws out.

\begin{figure}
    \centering
    \includegraphics[width=12cm]{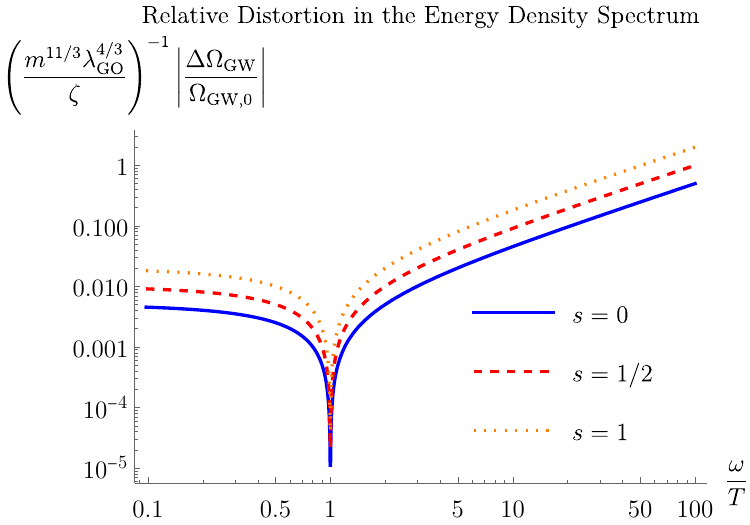}
    \caption{Plot of the absolute value of the relative distortion in the primordial energy density spectrum due to a dark matter candidate of spin $s\in\{0, \frac{1}{2}, 1\}$ at redshift $z$. The initial spectrum is assumed to follow a power law $\Omega_{\rm GW, 0} \sim (\omega/T)^2$. The region $\omega/T < 1$ corresponds to where $\Delta\Omega_{\rm GW}$ is negative. We've factored out the explicit dependencies that the primordial energy spectrum has on the mass and cutoff wavelength to show how the distortion scales with both parameters below the Planck scale. The parameter, $\zeta$ is composed of various constants given in Eqn. (\ref{zeta_constant}).}
    \label{distortion}
\end{figure}

Our last step is to relate the distortion in the primordial occupancy number to the primordial energy density spectrum
\begin{equation}
    \dfrac{\Delta\Omega_{\rm GW}(x,z;s)}{\Omega_{\rm GW, 0}(x,z)} = \dfrac{\Delta n(x,z;s)}{n_0(x,z)},
\end{equation}
where $\Delta\Omega_{\rm GW}(x,z;s)$ is the deviation from the primordial energy density spectrum as a function of the dimensionless frequency $x$ and $n_0(x,z)$ is the initial occupation index before any scattering has taken place. We can also write out the explicit mass and cutoff wavelength dependence that the relative distortion has
\begin{equation}
    \dfrac{\Delta\Omega_{\rm GW}(x,z;s)}{\Omega_{\rm GW, 0}(x,z)} = \frac{12\pi(G^5m^{11}\lambda^4_{\rm GO})^{1/3}H_0T\Omega_\chi}{\Omega_M^{1/2}}g(x,z;s) = \frac{m^{11/3}\lambda^{4/3}_{\rm GO}}{\zeta}g(x,z;s),
\end{equation}
where $g(x,z;s)$ contains all the dependencies that the relative distortion has on the dimensionless frequency $x$, redshift $z$, and spin $s$ with the new constant $\zeta$ is given by
\begin{equation}\label{zeta_constant}
    \zeta \equiv \frac{\Omega_M^{1/2}}{12\pi G^{5/3}H_0T\Omega_\chi} = \frac{m}{T}\xi.
\end{equation}
Figure~\ref{distortion} shows the absolute value of the relative distortion in the dimensionless energy density spectrum of a cosmological background sourced from single-field inflation with an initial power law spectrum, $\Omega_{\rm GW, 0} \sim (\omega/T)^2$, for DM candidates of different spins. Recall that on the CMB, the frequency for that determines Compton upscatter and downscatter is called the null. In our case the null frequency is $\omega/T \simeq 1$. We see that below the null (i.e. $\omega/T < 1$), the spectrum is negative, which is indicative of a loss of power/energy at lower frequencies, and the amplitude increases above the null (i.e. $\omega/T > 1$). We can understand this from the analogous effect on the CMB. Since particle number is conserved, the deficit of power is a result of a loss of gravitons/gravitational waves at lower frequencies. This is due to the elastic scattering between gravitons and the DM candidate where the kinetic energy of the DM gets transferred to the gravitons. This leads to a corresponding surplus of gravitons at higher frequencies, which thus causes an overall shifting of the spectrum at these higher frequencies. The fact that higher spin particles induce a greater distortion also has a simple interpretation: particles with spin $s > 0$ have both translational \emph{and} rotational kinetic energy. Therefore the higher spin particles have more energy that the gravitons are able to extract from a single collision.

From here, the thoughtful reader might ask if there is a particular frequency band for which this effect of scattering is particularly emphasized. We have shown that the entire spectrum gets modified as a result of this scatttering and thus there can't be a particular frequency band where this effect is most pronounced. We can think of this from the case of the CMB where photons scatter off hot electrons in the atmosphere of galaxy clusters: since CMB photon energies, $E_\gamma$, are much smaller than the rest mass of electrons, the fractional change of energy is $\sim \order{E_\gamma/m_e}$. Thus, the photon energies will be preferentially boosted from this scattering resulting in the fractional change of the intensity of the CMB, $\Delta I/I$, to appear dimmer for photon frequencies that are smaller than the kinetic energy of the electron (i.e. its temperature) and brighter for the frequencies that are above the kinetic energy of the electron. 

An equivalent way to think of this is by saying there are less photons in the frequencies that lie below the temperature of the electron because those photons have stolen some of the kinetic energy and they now have higher energies corresponding to an increase of photons in the frequency band that's above the electron energy. Now on the CMB, there is a sense in which there is a particular frequency band where you would preferentially look to measure this effect, i.e. the peak of the spectrum. But this is because we know that the initial spectrum is a black body and we can pinpoint the frequency that separates Compton upscattering and downscattering (also called the null). These two signatures are key to identifying whether the original spectrum has been distorted. However, this is set by observational constraints and \emph{not} fundamental physics.

This is the central purpose and hence the key takeaway of this manuscript: we have demonstrated the exact same phenomenon of Compton upscatter and downscatter on CGWBs due to scattering with a toy model of DM. The observational constraint would be much harder since the initial spectrum is a power law and therefore there is no obvious feature to look for, but it's fundamentally the same physics. The energy shift experienced by the gravitons is small and therefore the rest of the spectrum gets modified slightly: we get Compton downscatter below the null frequency connected to the kinetic energy of the dark matter, and we have Compton upscatter above the null frequency, so the question of ``which frequency band is this effect most emphasized" is ill-defined in this particular circumstance.

\subsection{Non-Thermal Particle Dark Matter}

Given the minuscule distortion for fundamental particle DM, one can ask what possible ways would allow for a larger distortion. Since the signal we have found is mass-dependent, considering higher mass candidates will naturally produce a greater distortion. This comes with the caveat that if one wants to study DM interactions with any cosmological gravitational wave background, because of the unitarity bounds, those DM candidates would need to be non-thermally produced. There are many examples of this, but one must contend with lifting the restriction of thermal equilibrium. 

One non-thermally produced DM candidate are primordial black holes (PBHs). For PBHs with cutoff wavelength $\lambda_{\rm GO} = 0.1\pc$ at a redshift $z = 1$, one can show that the optical depth scales like $\tau_{\rm PBH} \sim \beta^{-1/3}10^{-6}$ where $\beta = M/M_\odot$ and $M$ is the mass of the PBH. Current constraints show that PBHs can still comprise an $\order{1}$ fraction of the DM if their mass is within the range of $M \sim 10^{-15} - 10^{-12} M_\odot$ \cite{Carr10, Carr16, Carr21, Green21}. One can get an estimate for a range of optical depths, $\tau_{\rm PBH} \sim 10^{-1} - 10^{-2}$. The characteristic velocity for stellar mass objects on a circular orbit in a galaxy is $v_c = 200\km\s^{-1}/(1 + z)$ \cite{Binney87}. Therefore, the largest Compton $y$-parameter\footnote{One might be tempted to believe that the largest signal occurs at high redshifts. Especially because the optical depth becomes large $\tau_{\rm PBH} \gg 1$ and independent of redshift in a radiation-dominated universe. However the velocity reduces immensely $v_c \sim (1 + z)^{-1}$ which results in much less kinetic energy to be transferred.} on a CGWB is $y \sim 5.5\times 10^{-8}$. The energy transfer coefficients, $J_\ell$, are $\order{10^4}$ thus $\Delta\Omega_{\rm GW} \sim 10^{-4}\Omega_{\rm GW,0},\ 10^{-1}\Omega_{\rm GW, 0}$ for $\omega \approx T_H$ where $T_H$ is the Hawking temperature of the PBH. This will be investigated in more detail in a subsequent work.

\section{Discussion and Future Outlook} \label{discuss}

We have derived a formalism for describing spectral distortions on the energy density spectrum of cosmological gravitational wave backgrounds due to Compton scattering with thermally produced, fundamental particle DM. We have shown that, for frequencies below the kinetic energy of the DM, there is a decrease in power due to the diminished number of gravitons at those frequencies with a corresponding increase in power for frequencies above the kinetic energy. This means the previously low-energy gravitons have been kicked up to higher energies as a result of elastic scattering. This is exactly analogous to the SZ effect on the CMB from inverse Compton scatter of hot electrons. The main difference is that the new spectrum is a broken power law, but the signal on any CGWB will be much smaller relative to what one finds for the CMB.

There are three reasons which explain the diminutive signal on the CGWB compared to that of the CMB. The chief reason is the fact that the gravitational interaction between fundamental particles with masses below the Planck scale is very weak, i.e. the coupling constant for gravity $Gm$ is much smaller than E\&M $\alpha \simeq 1/137$. As a result, we would generically expect a tiny distortion. The second reason can be attributed to the fact that we are looking at cold DM. The temperature of electrons in a typical cluster is $T_e \simeq 10^8\K$ \cite{Birkinshaw99}, compared to the estimated temperature of cold DM of $T \simeq 2.7\K$. This means the amount of energy that gravitons can extract from the DM through elastic scattering is substantially smaller than what photons extract from electrons. Lastly, the relative abundance of the massive target also makes a difference. The number density of electrons in a cluster is $n_e \simeq 3\times 10^{-3}\cm^{-3}$, whereas for a DM candidate with $m = 10\eV[T]$, the number density is $n_\chi \simeq 3\times 10^{-5}\cm^{-3}$. All of these differences contribute to drive down the strength of the signal.

Other future work is incorporating higher-spin DM into the analysis of the gravitational SZ effect. Indeed, studies on higher-spin DM and possible detection mechanisms have been done \cite{Alexander21, Falkowski21, Banerjee22, Jenks22}. Introducing relativistic corrections could also be a fruitful venture since one can extend many of these arguments to larger redshifts. Moreover, we focused on the tensor perturbations generated by single-field inflation for the purpose of studying the early-Universe behavior of DM, but GWs from any source can experience elastic scattering. Therefore, one should expect similar distortions on astrophysical gravitational wave backgrounds. 

Additionally, there is still the issue of the angular singularity for which we needed to implement a cutoff angle. This causes the final answer to be dependent on that reference angle, which is undesirable. Over the decades, techniques to renormalize these sorts of divergences have been developed, hence a valuable follow-up work would be to extend those methods to the case of gravitational scattering. Some progress has been recently made in this endeavor. In Ref.~\cite{Gonzo23}, the 1-loop correction to the gravitational Compton scattering cross section was computed. By summing up the virtual contributions from 72 diagrams, they were able to show that the divergence cancels out.

One can also wonder about the angular scales for which this scattering will take place. For the analogous case in the CMB, since the scattering typically occurs in the atmosphere of galaxy clusters, the angular resolution that's needed corresponds to how well one's telescope is able to resolve individual clusters. This is typically $\lesssim$ 1 arcminute for clusters (see Ref.~\cite{Carlstrom02} for the full details). Because we are considering a cosmological distribution of DM in the early universe, the angular resolution that is required is quite small, commensurate with the very tiny signal that we get for our toy model of DM.

Lastly, it is important to note that this formalism is general enough to allow for the complete replacement of gravitational interactions by considering an entirely different background. One potentially promising new direction is studying Compton scattering of a neutrino background. Because neutrinos decouple when the universe was $T_\nu \sim 1\eV[M]$, \cite{Kolb90} they can also provide a wealth of knowledge of the universe, similar to a cosmic gravitational or microwave background. However, the prospects for detecting this background are significantly lower compared to any CGWB.

\section*{Acknowledgements}

The authors would like to thank Cyril Creque-Sarbinowski, David Kaiser, and Arthur Kosowsky for reviewing this manuscript and providing substantial comments about the work. M.H. would like to thank Michael Saavedra and Ira Rothstein for very enlightening conversations and for the suggestion of utilizing spinor-helicity variables, as well as explaining the origin of the forward scattering pole. M.H. would also like to thank Andrew Zenter, Ayres Freitas, and Brian Batell for fruitful discussions and patiently answering questions. Diagrams were drawn using \emph{TikZFeynman} \cite{Ellis17} and plots were made using MaTeX \cite{Horvat22}. We also made use of helpful unit conversion tables found in \cite{Tomberg21}. This project has been partially completed at the Laboratory for Nuclear Science and the Center for Theoretical Physics at the Massachusetts Institute of Technology (MIT-CTP/5492).
\newpage
\appendix

\section{Cosmological Gravitational Wave Background Formalism} \label{CGWB form}

We follow the treatment of cosmological backgrounds as described in \cite{Maggiore07}. Cosmological backgrounds of GWs can emerge from the incoherent superposition of a large number of astrophysical sources that are too weak to be detected separately and such that the number of sources that contribute to each frequency bin is much larger than one. We make some assumptions about the background.

\paragraph{Stationarity} This means that all n-point correlation functions can only depend on time differences as opposed to absolute time, i.e. $$\expval{h_A(t)h_B(t')} \propto f(t - t'),$$ but not on $t,t'$ separate. As a result, we must have $$\expval{\tilde{h}^*_A(\nu)\tilde{h}_B(\nu')} \propto \delta(\nu - \nu').$$ Thus, the typical time scale it can change substantially is of order the age of the Universe.

\paragraph{Gaussianity} All n-point correlators are or can be reduced to sums and products of the 2-point correlation function (and the vacuum expectation value but since we impose stationarity, the vacuum expectation value has to be a constant that we set to zero for simplicity). This is a direct consequence of the central limit theorem.

\paragraph{Isotropy} Because the early Universe was highly isotropic (and we know this from the CMB), we expect the gravitational background should be isotropic as well. This implies $$\expval{\tilde{h}^*_A(\nu,\vu{n})\tilde{h}_B(\nu',\vu{n}')} \propto \delta^2(\vu{n},\vu{n}'),$$ where
\begin{equation}
    \delta^2(\vu{n},\vu{n}') \equiv \delta(\cos\theta - \cos\theta')\delta(\phi - \phi').
\end{equation}
This comes from the idea that waves coming from different directions should be uncorrelated.

\paragraph{Polarization} Lastly, we expect the background to be unpolarized, i.e. $$\expval{\tilde{h}^*_A(\nu,\vu{n})\tilde{h}_B(\nu',\vu{n}')} \propto \delta_{AB}.$$

All of these conditions taken together gives us
\begin{equation}
    \expval{\tilde{h}^*_A(\nu,\vu{n})\tilde{h}_B(\nu',\vu{n}')} = \delta(\nu - \nu')\frac{\delta^2(\vu{n},\vu{n}')}{4\pi}\delta_{AB}\frac{S_h(\nu)}{2},
\end{equation}
where $S_h(\nu)$ is the spectral density of cosmological backgrounds with dimensions $\Hz^{-1}$. $S_h(\nu)$ is an even function, and the factor of 4$\pi$ is there for normalization purposes. While the spectral density represents the measurable quantity, it is not \emph{a priori} clear how it connects to the physics. Thankfully, we can connect this quantity to the frequency-varying spectral density for the fractional energy density $\Omega_{\rm GW}(\nu)$
\begin{equation}
    \Omega_{\rm GW}(\nu) = \frac{1}{\rho_c}\dv{\rho_{\rm GW}}{\log\nu} = \frac{(2\pi)^2}{3H_0^2}\nu^3S_h(\nu),
\end{equation}
where $\rho_c$ is the critical energy density required to close the Universe, which is given by $\rho_c = 3H_0^2/(8\pi G) \simeq 1.688\times 10^{-8}h^2\erg/\cm^3$, $H_0$ is the present-day Hubble parameter given by $H_0 = 100h\km\s^{-1}\pc[M]^{-1}$, with $h$ being the little Hubble parameter which is used to parametrize the uncertainty of the Hubble parameter. We define $h \simeq 0.70$ and the frequency-varying spectral energy density is related to the (mean) energy density of GWs by 
\begin{equation}
    \rho_{\rm GW} = \frac{1}{32\pi G}\expval{\dot{h}^{ij}\dot{h}_{ij}} = \int_{\nu = 0}^{\nu = \infty}\dd{(\log\nu)}\dv{\rho_{\rm GW}}{(\log\nu)}.
\end{equation}
It will be prudent to work with $h_0^2\Omega_{\rm GW}(\nu)$ as a way to circumvent any potential uncertainty. Because we are interested in the derivation for the gravitational analog of the SZ effect on cosmological backgrounds, an important quantity is the occupation number of gravitons per cell of phase space $n(\br,\bp) = n_\nu$, which only depends on the frequency. The frequency is related to the momentum by $|\bp| = \omega = 2\pi\nu$, and the fact that it only depends on the magnitude and not direction is a consequence of the isotropy condition we placed on cosmological backgrounds. The energy density is related to the number density by
\begin{equation}
    \rho_{\rm GW} = 2\int\dfrac{\dd[3]{p}}{(2\pi)^3}\omega(\bp)n_\nu,
\end{equation}
where the factor of 2 in front is for the two polarizations of the graviton. Thus the spectral density is
\begin{equation}
    \dv{\rho_{\rm GW}}{\log\nu} = (4\pi\nu^2)^2n_\nu\Rightarrow h^2\Omega_{\rm GW}(\nu) = \frac{8\pi G}{3} \frac{h^2}{H^2_0}(4\pi\nu^2)^2n_\nu.
\end{equation}

\section{Spinor-Helicity Formalism} \label{spin-hel}

The spinor-helicity formalism is a modern technique for the calculation of scattering amplitudes by enabling much faster computation through the use of Weyl spinor algebra. This formalism starts from the recognition that $SO(1,3)$ and $SL(2,\mathbb{C})$ are isomorphic as vector spaces, and therefore any element in one of these sets has an associated representation in the other. In this case, we start off with a 4-momentum $p^\mu \in SO(1,3) \mapsto p_\mu\sigma^\mu_{a\dot{b}} \equiv p_{a\dot{b}} \in SL(2,\mathbb{C})$ where $\sigma^\mu_{a\dot{b}} = (I_{a\dot{b}},\sigma^i_{a\dot{b}})$, and we call $p_{a\dot{b}}$ a bi-spinor. When $p^\mu = (E,\bp)$ is the momentum of a massless particle, we can write the null vector as a product of two Weyl spinors
\begin{equation}
    P_{a\dot{b}} = -\lambda_a\tilde{\lambda}_{\dot{b}},
\end{equation}
where
\begin{equation}
    \lambda_a = \begin{bmatrix}
	\sqrt{E + p^z} \\ \frac{p^x + ip^y}{\sqrt{E + p^z}}
	\end{bmatrix}, \hspace{0.5cm}\tilde\lambda_{\dot{b}} = \begin{bmatrix}
	\sqrt{E + p^z} \\ \frac{p^x - ip^y}{\sqrt{E  + p^z}}
	\end{bmatrix}.
\end{equation}
Next, we introduce the notation $\bar\sigma^{\mu\dot{a}b} = (I_2, -\sigma^i)^{\dot{a}b} \equiv \ep^{bc}\ep^{\dot{a}\dot{d}}\sigma^\mu_{c\dot{d}}$, where we raise and lower the spinorial indices using the Levi-Civita symbol. The 2-component massless Weyl spinors are often denoted by $\lambda_a$, $\tilde{\lambda}_{\dot{a}}$ and are usually written using the spinor bra-ket notation
\begin{equation}
    \lambda_a = |p]_a \equiv |p],\hspace{0.5cm}\tilde{\lambda}^{\dot{a}} \equiv \ket{p}^{\dot{a}} \equiv \ket{p},\hspace{0.5cm }\lambda^a = [p|^a \equiv [p|, \hspace{0.5cm}\tilde{\lambda}_{\dot{a}} \equiv \bra{p}_{\dot{a}} \equiv \bra{p},
\end{equation}
where the indices $a, b, \ldots$ are the $SL(2, \mathbb{C})$ matrix indices. The momentum can then be written as a product of these 2-component spinors 
\begin{equation}
    p_{a\dot{b}} = -\lambda_a\tilde{\lambda}_{\dot{b}} = -|p]_a\bra{p}_{\dot{b}}, \hspace{0.5cm} p^{\dot{a}b} = -\tilde{\lambda}^{\dot{a}}\lambda^b = -\ket{p}^{\dot{a}}[p|^b.
\end{equation}
A consequence of these being massless spinors is that they satisfy the massless Dirac equation, which in these new variables is given in the form
\begin{equation}
    p_{a\dot{b}}\ket{p}^{\dot{b}} = 0,\hspace{0.5cm}\bra{p}_{\dot{a}}p_{\dot{a}b} = 0,\hspace{0.5cm}[p|^a p_{a\dot{b}} = 0, \hspace{0.5cm}p^{\dot{a}b}[p|_b = 0.
\end{equation}
The same treatment can be extended to massive spinors - originally written in \cite{Arkani-Hamed17} and expounded upon by \cite{Ochirov18} - by introducing $SU(2)$ indices $I, J,\ldots$ to the Weyl spinors. We can represent the momenta for massive particles to be
\begin{equation}
    p_{a\dot{b}} = \lambda^I_a\tilde{\lambda}_{\dot{b} I} \equiv \lambda^I_a\tilde{\lambda}^J_{\dot{b}}\ep_{IJ}, \hspace{0.5cm}p^{a\dot{b}} = \tilde{\lambda}^{aI}\lambda^{\dot{b}}_I \equiv \tilde{\lambda}^{aI}\lambda^{\dot{b}J}\ep_{IJ},
\end{equation}
where the Levi-Civita tensor $\ep_{IJ}$ acts as a metric on the space of $SU(2)$ spinors. These momenta come equipped with their own bra-ket notation in the form of
\begin{equation}
    p_{a\dot{b}} = \ket{p^I}_a[p_I|_{\dot{b}},\hspace{0.25cm} p^{\dot{a}b} = -|p^I]^{\dot{a}}\bra{p_I}^b,\hspace{0.25cm} \ket{p^I}_a\bra{p_I}^{b} = -m\delta^b_a, \hspace{0.25cm} |p^I]^{\dot{b}}[p_I|_{\dot{a}} = m\delta^{\dot{b}}_{\dot{a}}.
\end{equation}
These momenta also follow the Dirac equation for massive particles
\begin{equation}
    p_{a\dot{b}}|p^I]^{\dot{b}} = m\ket{p^I}_a, \hspace{0.5cm}\bra{p^I}^a p_{a\dot{b}} = [p^I|_{\dot{b}}.
\end{equation}
It is also the convention to use the following shorthand notation when referencing these bi-spinor momenta
\begin{equation}
    \ket{p_i} \equiv \ket{i}, \hspace{0.5cm}|p_j] \equiv |j],
\end{equation}
and so we shall do the same, where we distinguish massless spinors from massive spinors by the placement of $SU(2)$ indices inside the bra-ket.\footnote{In order for the massive spinors to transform covariantly, spinor products must be a symmetrized combination of normal spinor products, so $\expval{1^I2^J}^2 \equiv \expval{1^I2^J}\expval{1^K2^L} + \expval{1^I2^L}\expval{1^K2^J}$ and $\expval{1^I4}^2 \equiv \expval{1^I4}\expval{1^K4}$.} For example, we write\footnote{Note that it is common to suppress the $SU(2)$ indices in favor of bold-facing the corresponding momenta, so $\expval{1^I4} \leftrightarrow \expval{\mathbf{1}4}$ and $\expval{1^I2^J} \leftrightarrow \expval{\mathbf{12}}$, but we shall not do that here.}
\begin{equation}
    [pq] \equiv [p|^a |q]_a = -[qp] = (\expval{qp})^*,\hspace{0.5cm}\expval{pq}[pq] = 2p\cdot q.
\end{equation}
Another quantity associated with massless particles are polarization vectors $\ep^\mu_\pm(p;q)$ given by
\begin{equation}
    \ep^\mu_+(p;q) = -\frac{\bra{q}\gamma^\mu|p]}{\sqrt{2}\expval{qp}}, \hspace{0.5cm}\ep^\mu_-(p;q) = -\frac{\bra{p}\gamma^\mu|q]}{\sqrt{2}[qp]},
\end{equation}
where $\gamma^\mu$ are the $\gamma$-matrices which satisfy $\acomm{\gamma^\mu}{\gamma^\nu} = -2\eta^{\mu\nu}$ and in the Weyl-representation are given by
\begin{equation}
    \gamma^\mu = \begin{bmatrix}
	0 & \sigma^\mu_{a\dot{b}} \\ \bar\sigma^{\mu\dot{a}b} & 0
	\end{bmatrix}.
\end{equation}
We also have the property that 
\begin{equation}
    [\ep^\mu_+(p;q)]^* = -\ep^\mu_-(p;q),
\end{equation}
where $q$ is an auxiliary momentum that does not represent a physical quantity, but merely reflects the gauge freedom of the theory. 

Another indispensable tool in the arsenal of Spinor-Helicity variables is the use of \emph{the little group scaling}. The little group is essentially a subgroup whose action leaves a particular state invariant. In our case, for momentum bi-spinors written as $p_{a\dot{b}} = -\ket{p}_a[p|_{\dot{b}}$, if we scaled the spinors by some parameter $t$ in the following way: $\ket{p}\rightarrow t\ket{p}$, $|p]\rightarrow t^{-1}|p]$, then the bi-spinor is left the same. If/when we allow momenta to be complex, then $t \in \mathbb{C}$ will also be allowed to be complex.
We can immediately deduce what the final form of the 3-point amplitude for three massless particles with momenta and helicity $(p_i,h_i)$ has to be up to an overall constant:
\begin{equation}
    \amp_3(1^{h_1}2^{h_2}3^{h_3}) = C\expval{12}^{h_3 - h_2 - h_1}\expval{13}^{h_2 - h_1 - h_3}\expval{23}^{h_1 - h_2 - h_3},
\end{equation}
which we write when $h_1 + h_2 + h_3 < 0$ and
\begin{equation}
    \amp_3(1^{h_1}2^{h_2}3^{h_3}) = \frac{C}{[12]^{h_3 - h_2 - h_1}[13]^{h_2 - h_3 - h_1}[23]^{h_1 - h_2 - h_3}},
\end{equation}
whenever $h_1 + h_2 + h_3 > 0$, and the constant $C$ is decided via dimensional analysis. Using these new helicity methods, it is possible to treat both sets of particles being on-shell at the cost of allowing the momenta to be complex. To deal with that, we introduce a \emph{shifted} momentum $\hat{p}_{ab} = p_{ab} + \Delta p_{ab}$, and we will enforce the on-shell condition on the shifted momentum where we allow $\hat{p}_{ab}$ to be complex. This extension to complex momenta guarantees the following: (1) momentum conservation for external particles, (2) both $\hat{p}_i$ and $\hat{p}_j$ are null vectors, and (3) the poles associated with propagators are simple. These conditions force an explicit form for the shifted complex momenta
\begin{equation}
    \hat{p}_i = p_i + z\eta \Leftrightarrow |\hat{\imath}] = |i] + z|j], \hspace{0.5cm}\ket{\hat{\jmath}} = \ket{j} - z\ket{i},\hspace{0.5cm}\ket{\hat{\imath}} = \ket{i}, \hspace{0.5cm}|\hat{\jmath}] = |j].
\end{equation}
The parameter $z$ acts as a pole for the associated amplitude. As a result, the shifted n-point amplitude $\hat{\amp}_n(z)$ that is obtained from shifting the momenta of particles $i$ and $j$ represents a simple pole at $z = z_{ib}$, which brings the job of calculating n-point amplitudes to essentially one of computing the residues, i.e.
\begin{equation}
    \Res[\hat{\amp}_n;z_{ib}] \equiv \lim\limits_{z\rightarrow z_{ib}}(z - z_{ib})\hat{\amp}_n(z).
\end{equation}
Let us illustrate this more explicitly by invoking the Cauchy integral formula
\begin{equation}
    B_n = \frac{1}{2\pi i}\oint_\gamma\dd{z}f(z),
\end{equation}
where $\gamma$ is some curve that encloses all the singularities of the function $f(z) \equiv \hat{\amp}_n(z)/z$ with $\hat{\amp}_n(0) = \amp_n$ and we regard $B_n$ as a "boundary term." The Residue theorem states that
\begin{equation}
    B_n = \sum_{\rm residues}\Res[f,\rm residues] = \Res[\frac{\hat{\amp}_n(z)}{z};0] + \sum_{z = z_{ib}}\Res[\frac{\hat{\amp}_n(z)}{z};z_{ib}].
\end{equation}
It follows that $\Res[\frac{\hat{\amp}_n(z)}{z};0] = \amp_n$ which leads to us writing
\begin{equation}
    \amp_n = -\sum_{z = z_{ib}}\Res[\frac{\hat{\amp}_n(z)}{z};z_{ib}] + B_n.
\end{equation}
In most situations of interest the boundary term will be zero. Lastly, we can use the fact that residues can be written as a product of on-shell lower-point amplitudes evaluated with complex momenta. This comes from the fact that the residue of a given pole at $z = z_{ib}$ comes from a Feynman diagram with two particles: $i$ and $b \neq j$ on one side of the propagator. As a result, the momentum that flows into the propagator is on-shell. What makes them on-shell is the fact that we've evaluated the lower-point amplitudes at $z = z_{ib}$. This means we can write
\begin{equation}
    \sum_{z = z_{ib}}\Res[\frac{\hat{\amp}_n(z)}{z};z_{ib}] = -i\amp_L(z_{ib})\frac{1}{P^2_{ib}}\amp_R(z_{ib}),
\end{equation}
where $\amp_L(z_{ib})$ is the on-shell amplitude for the particles $i,b\neq j$ and the one in the propagator, and $\amp_R(z_{ib})$ is the amplitude for the particle in the propagator and the remaining particles in the n-point amplitude. This can be generalized to amplitudes with multiple simple poles as well as particles with different helicities
\begin{equation}
    \amp_n = \sum_{z_{ib}}\sum_{h}i\amp_L(z_{ib};h)\frac{1}{(P^h_{ib})^2}\amp_R(z_{ib};h),
\end{equation}
and similar arguments can be made when the propagator is massive.

The next piece of machinery we will utilize comes from string theory. The Kawai-Lewellen-Tye (KLT) relations is a reduction of closed-string amplitudes to sums and products of open-string amplitudes \cite{Kawai85, Bern98}. In the field theory limit, this reduction forces the tree-level amplitude for gravitational scattering to factorize into products of amplitudes of Yang-Mills tree-level amplitudes\footnote{It was also found in \cite{Choi95} that the gauge and Lorentz invariance of linearized general relativity, when coupled to matter fields, also leads to a factorization of the scattering amplitudes.} \cite{Bohr14, Bohr16, Holstein17}. Thus, we can express the tree-level amplitude between a graviton and an arbitrary spin particle in the following way
\begin{equation}
    i\mcal_{s}(p_i, p_f, k_i, k_f) = \frac{\kap^2}{e^4}\qty(\frac{p_i\cdot k_i p_i\cdot k_f}{k_i\cdot k_f})\amp_s(p_i, k_f, p_f, k_i)\amp_{s = 0}(p_i, k_f, p_f, k_i),
\end{equation}
where $\amp_{s = 0}$ is the Compton scattering amplitude for a photon scattering off of a scalar particle and $\amp_s$ is the amplitude for a photon to scatter off a particle of spin $s \in \{0, \frac{1}{2}, 1\}$. Here the indices $i, f$ denote the initial and final momenta are the graviton and the target dark matter particle. 

The color-stripped ordered scattering amplitude derived in the helicity formalism for gluons with massive scalars \cite{Badger05}, gluons with massive spin-1/2 fermions \cite{Ochirov18}, and gluons with massive spin-1 vectors \cite{Ballav21} can all be written down in a general formula for gluons with the same or differing helicities. For gluons with the same helicity (which we denote with a + in the superscript over their momentum), the amplitude becomes
\begin{equation}
    \amp_s(p_i, p_f, k_i^+, k_f^+) = e^2\frac{(-1)^{s + 1}m^{2 - 2s}[k_i k_f]^2}{4k_i\cdot k_fp_i\cdot k_i}\expval{1^I2^J}^{2s}, 
\end{equation}
and for gluons with opposing helicities (taken from \cite{Arkani-Hamed17}), the amplitude can be written as
\begin{equation}
    \amp_s(p_i, p_f, k_f^+, k_i^-) = e^2\frac{\bra{k_f}p_i|k_i]^{2 - 2s}}{4k_i\cdot k_fp_i\cdot k_i}\qty(\expval{1^I3}[2^J4] + [1^I4]\expval{2^J3})^{2s}.
\end{equation}
Furthermore, there is an additional monodromy relation which brings the amplitudes to the form
\begin{equation}
    \amp_{s}(p_i,k_f^+,p_f,k_i^+) = \frac{k_i\cdot k_f}{p_i\cdot k_f}\amp_{s}(p_i,p_f,k_f^+,k_i^+) = e^2\frac{(-1)^{s + 1}m^{2 - 2s}[k_i k_f]^2}{4p_i\cdot k_i p_i\cdot k_f},
\end{equation}
and
\begin{equation}
    \amp_{s}(p_i,k_f^+,p_f,k_i^-) = \frac{k_i\cdot k_f}{p_i\cdot k_f}\amp_{s}(p_i,p_f,k_f^+,k_i^-) = e^2\frac{\bra{k_f}p_i|k_i]^{2 - 2s}}{4p_i\cdot k_ip_i\cdot k_f}\qty(\expval{1^I3}[2^J4] + [1^I4]\expval{2^J3})^{2s}.
\end{equation}
This brings the gravitational Compton scattering amplitude for gluons with the same helicities \cite{Chiodaroli22} to be
\begin{equation}
    \mcal_s(p_i,p_f,k_i^+,k_f^+) = (-1)^s\frac{\kap^2}{16}\frac{[k_i k_f]^4m^{4-2s}\expval{1^I2^J}^{2s}}{(k_i\cdot k_f) (p_i\cdot k_i) (p_i\cdot k_f)},
\end{equation}
and for opposite-helicity gluons we have
\begin{equation}
    \mcal_s(p_i,p_f,k_i^-,k_f^+) = -\frac{\kap^2}{16}\frac{\bra{k_f}p_i|k_i]^{4 - 2s}}{(k_i\cdot k_f)(p_i\cdot k_i)(p_i\cdot k_f)}\qty(\expval{1^I3}[2^J4] + [1^I4]\expval{2^J3})^{2s}.
\end{equation}
The other helicities are related by complex conjugation where $i\mcal_{s}(p_i, p_f, k_f^-, k_i^-) = \newline (i\mcal_{s}(p_i, p_f, k_f^+, k_i^+))^*$ and $i\mcal_{s}(p_i, p_f, k_f^-, k_i^+) = (i\mcal_{s}(p_i, p_f, k_f^+, k_i^-))^*$, and thus finding the cross section for the corresponding target particle is a matter of setting the $s$ to be some spin. It should be noted that for this article, we rewrote the 4-momenta in the following variables: $p_i \equiv p = (E, \bp)$, $k_i \equiv k = (\omega, \bk)$, $p_f \equiv p' = (E', \bp')$, $k_f \equiv k' = (\omega', \bk')$.

\section{The Optical Depth} \label{optical_depth}

Since most of the interactions of interest occurs during the radiation- and matter-dominated epochs of the Universe, we used an analytic expression of the scale factor that incorporates both eras given in \cite{Maggiore18}
\begin{equation}
    a(\eta) = a_{\rm eq}\qty[\frac{2\eta}{\eta_*} + \qty(\frac{\eta}{\eta_*})^2],
\end{equation}
where $\eta$ is again the conformal time while $a_{\rm eq}$ is the scale factor at matter and radiation equality and 
\begin{equation}
    \eta_* = \frac{2\Omega_R^\frac{1}{2}}{H_0\Omega_M}.
\end{equation}
This expression for the scale factor gives us a nice and convenient way to write the conformal time as a function of the scale factor in order to swap to redshift space
\begin{equation}
    \eta(z) = \frac{2}{H_0\sqrt{\Omega_M}}\qty[\sqrt{a(z) + a_{\rm eq}} - \sqrt{a_{\rm eq}}] = \frac{2}{H_0\sqrt{\Omega_M}}\qty[\sqrt{\frac{2 + z_{\rm eq} + z}{(1 + z)(1 + z_{\rm eq})}} - \frac{1}{\sqrt{1 + z_{\rm eq}}}],
\end{equation}
with $z_{\rm eq} = \frac{\Omega_M}{\Omega_R} - 1$ being the redshift of matter-radiation equality. For a model of cosmological DM, because we are mainly interested in DM at the epoch in which it freezes out, we can write
\begin{equation}
    n_\chi(z) = \frac{\rho_\chi(z)}{m} = \frac{\rho_{c}}{m}\Omega_\chi(z) = \frac{\rho_{c}\Omega_\chi}{m}(1 + z)^3,
\end{equation}
where we wrote $\rho_\chi = \rho_{c}\Omega_\chi$. The rest-frame cross section for each spin is
\begin{equation}
    \sigma_{s = 0}(m, \omega, \lambda) = 2\pi(Gm)^2\int_{\theta_{\min}(\lambda)}^{\theta_{\max}}\dd{\theta}\qty(\frac{\omega'(\theta)}{\omega})^2\qty[\cot[4](\frac{\theta}{2})\cos[4](\frac{\theta}{2}) + \sin[4](\frac{\theta}{2})]
\end{equation}
for graviton-scalar interactions,
\begin{equation}
    \begin{split}
        \sigma_{s = 1/2}(m, \omega, \lambda) =&\ 2\pi(Gm)^2\int_{\theta_{\min}(\lambda)}^{\theta_{\max}}\dd{\theta}\qty(\frac{\omega'(\theta)}{\omega})^3\left[\cot[4](\frac{\theta}{2})\cos[4](\frac{\theta}{2}) + \sin[4](\frac{\theta}{2})\right.\\ &\left.+ \frac{2\omega}{m}\qty(\cot[2](\frac{\theta}{2})\cos[6](\frac{\theta}{2}) + \sin[6](\frac{\theta}{2})) + 2\qty(\frac{\omega}{m})^2\qty(\cos[6](\frac{\theta}{2}) + \sin[6](\frac{\theta}{2}))\right]
    \end{split}
\end{equation}
for graviton-fermion scattering, and
\begin{equation}
    \begin{split}
        \sigma_{s = 1}(m, \omega, \lambda) =&\  2\pi(Gm)^2\int_{\theta_{\min}(\lambda)}^{\theta_{\max}}\dd{\theta}\qty(\frac{\omega'(\theta)}{\omega})^4\left[\qty(\cot[4](\frac{\theta}{2})\cos[4](\frac{\theta}{2}) + \sin[4](\frac{\theta}{2}))\qty(1 + \frac{2\omega\sin[2](\frac{\theta}{2})}{m})^2\right.\\ &\left. +\frac{16}{3}\frac{\omega^2}{m^2}\qty(\cos[6](\frac{\theta}{2}) + \sin[6](\frac{\theta}{2}))\qty(1 + \frac{2\omega\sin[2](\frac{\theta}{2})}{m}) \right.\\&\left.+ \frac{16}{3}\frac{\omega^4}{m^4}\sin[2](\frac{\theta}{2})\qty(\cos[4](\frac{\theta}{2}) + \sin[4](\frac{\theta}{2}))\right]
    \end{split}
\end{equation}
for graviton-massive vector bosons, where
\begin{equation}
    \frac{\omega'(\theta)}{\omega} = \frac{1}{1 + \frac{2\omega}{m}\sin[2](\frac{\theta}{2})} \simeq 1,
\end{equation}
in the low energy limit. The optical depth can then be defined by
\begin{equation}
    \dv{\tau}{t} = n_\chi\sigma_s\Rightarrow \tau = \int^{t_{\rm scat}}_{t_O} n_\chi\sigma_s\dd{t} = \int^{\eta_{\rm scat}}_{\eta_O}\dd{\eta}n_{\chi}(\eta)\sigma_s(m, \lambda, \eta)a(\eta)
\end{equation}

\section{Energy Transfer Integral Expressions} \label{energy_trans}

Here we explore some of the interesting properties of the energy transfer integrals found by expanding $\qty(1 - \bp\cdot\vu{n}/m)\Delta\dv*{\sigma_s}{\Omega}$ to second order in $\bp/m$ and performing Gaussian integrals. The expression for energy transfer of scalar field DM and gravitons is
\begin{equation}
    \begin{split}
        J_1(x,\lambda;0) &= \frac{16\sigma_0n_\chi T}{m}x(4 - x)\qty[2\pi\int^{\beta(\lambda)}_{-1}\frac{1 + 6\cos^2\theta + \cos^4\theta}{1 - \cos\theta}\dd{\cos\theta}],\\ \hspace{0.5cm} J_2(x,\lambda;0) &= \frac{16\sigma_0n_\chi T}{m}2x^2\qty[2\pi\int^{\beta(\lambda)}_{-1}\frac{1 + 6\cos^2\theta + \cos^4\theta}{1 - \cos\theta}\dd{\cos\theta}].
    \end{split}
\end{equation}
This particular form is intriguing since the angular integral is nothing but $1 - \cos\theta \times$ the rest-frame low-energy angular functions for the gravitational Compton differential cross section. The polynomials are also exactly identical to the ones found for the E\&M case, i.e.
\begin{equation}
    \begin{split}
        I_1(x) &= \frac{\sigma_Tn_eT}{m}x(4 - x)\qty[2\pi\int_{-1}^{1}(1 - \cos\theta)(1 + \cos^2\theta)\dd{\cos\theta}], \\ I_2(x) &= \frac{\sigma_Tn_eT}{m}2x^2\qty[2\pi\int_{-1}^{1}(1 - \cos\theta)(1 + \cos^2\theta)\dd{\cos\theta}],
    \end{split}
\end{equation}
where $\sigma_T$ is the Thomson cross section and $n_e$ is the number density of electrons. This seems to suggest a universal behavior for Compton scattering of radiation off of non-relativistic matter in a thermal bath:
\begin{equation}
    \begin{split}
        J_1(x,\lambda;s) &= \frac{8\sigma nT}{m}x(4 - x)\qty[2\pi\int_{\theta_{\min}}^{\theta_{\max}}(1 - \cos\theta)\Theta_s(\theta)\dd{\theta}],\\ J_2(x,\lambda;s) &= \frac{8\sigma nT}{m}2x^2 \qty[2\pi\int_{\theta_{\min}}^{\theta_{\max}}(1 - \cos\theta)\Theta_s(\theta)\dd{\theta}],
    \end{split}
\end{equation}
where $\Theta_s(\theta)$ are the angular functions for either gravity or E\&M with $s$ being the matter target. For spin-1/2 and spin-1 particles, this universal behavior takes on the form
\begin{equation}
    \begin{split}
		J_1\qty(x,\lambda;\frac{1}{2}) &= \frac{16\sigma_\frac{1}{2}n_\chi T}{m}x(4 - x)\qty[2\pi\int^{\beta(\lambda)}_{-1}\frac{(1 + \cos\theta)[(1 - \cos\theta)^4 + (1 + \cos\theta)^3]}{1 - \cos\theta}\dd{\cos\theta}],\\ J_2\qty(x,\lambda;\frac{1}{2}) &= \frac{16\sigma_\frac{1}{2}n_\chi T}{m}2x^2 \qty[2\pi\int^{\beta(\lambda)}_{-1}\frac{(1 + \cos\theta)[(1 - \cos\theta)^4 + (1 + \cos\theta)^3]}{1 - \cos\theta}\dd{\cos\theta}],
    \end{split}
\end{equation}
and 
\begin{equation}
    \begin{split}
		J_1\qty(x,\lambda;1) &= \frac{32\sigma_1n_\chi T}{m}x(4 - x)\qty[2\pi\int^{\beta(\lambda)}_{-1}\frac{(1 + \cos\theta)^2[3(1 - \cos\theta)^4 + (1 + \cos\theta)^2]}{1 - \cos\theta}\dd{\cos\theta}],\\ J_2\qty(x,\lambda;1) &= \frac{32\sigma_1n_\chi T}{m}2x^2 \qty[2\pi\int^{\beta(\lambda)}_{-1}\frac{(1 + \cos\theta)^2[3(1 - \cos\theta)^4 + (1 + \cos\theta)^2]}{1 - \cos\theta}\dd{\cos\theta}].
    \end{split}
\end{equation}
This suggests we can write the energy transfer coefficients parameterized as a following function of spin:
\begin{equation}
    \begin{split}
        J_1\qty(x,\lambda;s) &= \frac{2^{2s + 3}\sigma_sn_\chi T}{m}\qty[2\pi\int^{\cos\theta_{\min}(\lambda)}_{-1}\frac{(1 + \cos\theta)^{2s}[a_s(1 - \cos\theta)^4 + (1 + \cos\theta)^{4 - 2s}]}{1 - \cos\theta}\dd{\cos\theta}]x(4 - x),\\ J_2\qty(x,\lambda;s) &= \frac{2^{2s + 3}\sigma_sn_\chi T}{m}\qty[2\pi\int^{\cos\theta_{\min}(\lambda)}_{-1}\frac{(1 + \cos\theta)^{2s}[a_s(1 - \cos\theta)^4 + (1 + \cos\theta)^{4 - 2s}]}{1 - \cos\theta}\dd{\cos\theta}]2x^2,
    \end{split}
\end{equation}
where $a_s$ is a sequence satisfying 
\begin{equation}
    a_s = \begin{cases}
        1, & s = 0\\1, & s = \frac{1}{2}\\3, & s = 1
	   \end{cases}
\end{equation}
Thus, assuming a closed form analytic expression exists for the sequence ${a_s}$, one would only need to specify the rest-energy cross section for Compton scattering between a graviton and spin-$s$ particle, and then one could skip to (\ref{eval_komp}) without much work, so long as the massive field is in a thermal bath.

\bibliographystyle{JHEP}
\bibliography{gravitational_wave_background_compton}

\end{document}